# Theory and Algorithms for Pulse Signal Processing

Gabriel Nallathambi and Jose C. Principe

*Abstract*—The integrate and fire converter transforms an analog signal into train of biphasic pulses. The pulse train has information encoded in the timing and polarity of pulses. While it has been shown that any finite bandwidth analog signal can be reconstructed from these pulse trains with an error as small as desired, there is a need for fundamental signal processing techniques to operate directly on pulse trains without signal reconstruction. In this paper, the feasibility of performing online the signal processing operations of addition, multiplication, and convolution of analog signals using their pulses train representations is explored. Theoretical framework to perform signal processing with pulse trains imposing minimal restrictions is derived, and algorithms for online implementation of the operators are developed. Performance of the algorithms in processing simulated data is studied. An application of noise subtraction and representation of relevant features of interest in electrocardiogram signal is demonstrated with mean pulse rate less than 20 pulses per second.

## I. INTRODUCTION

One of the central principles in signal processing is the Whittaker-Shannon-Nyquist sampling theorem, which states that there is no loss of information between bandlimited analog signals and digital representations if the sampling rate is at least twice the maximum frequency present in the analog signal of interest [1]–[3]. Driven by sampling theory, programming flexibility and transistor scaling, nearly all data acquisition, processing and communication has progressed from continuous domain to the digital domain [4]. These advances along with the availability of high fidelity, low cost analog to digital converters (ADC) and digital signal processors (DSP) have led to an exponential increase in the digitalization of information processed from analog world sources [5]. The sampling theorem is a worst-case theorem, because it assumes that the highest frequency of input signal is always present, which normally is not the case. Conventional Nyquist sampling results in highly redundant sample representations that can overwhelm bandwidth in communications, and DSPs in real time portable applications [5]; therefore, efficient sensing and intelligent processing of sensor data for emerging applications requires new fundamental advances in the theory and implementation of data acquisition, conversion, and signal processing.

Recent developments in alternative sampling schemes such as compressive sensing [6], finite rate of innovation [7], and signal-dependent time-based samplers [8]–[10] are promising. These approaches combine sensing and compression into a single step by recognizing that useful information in real world signals is sparser than the raw data generated by sensors. The focus of this paper is on processing of pulse trains created by a special type of analog to pulse converter named integrate and fire converter (IFC), which converts an analog signal of finite bandwidth into a train of pulses where the area under the curve of the analog signal is encoded in the time difference between pulses [10].

The IFC is inspired by the leaky integrator and fire neuron model [11]. It takes advantage of the time structure of the input, enabling users to tune the IFC parameters for sensing specific regions of interest in the signal; therefore, it provides a compressed representation of the analog signal, using the charge time of the capacitor as the sparseness constraint [12]–[14]. Rastogi et al. [10] studied the hardware implementation of the IFC and showed that the power consumption and area required is smaller than most of the ADCs available: a single channel IFC has ~ 30 transistors with a figure of merit of 0.6 pJ/conv for an 8–bit converter, implemented using CMOS 0.6 $\mu m$ technology in a layout box of 100 $\mu m$ X 100 $\mu m$. Feichtinger et al. [15] proved mathematically the conditions for finite bandwidth analog signal to be approximately reconstructed from the train of IFC pulses with an error as small as desired. The simplicity in IFC sampling is balanced by complex non-linear reconstruction algorithm

Various processing schemes have been proposed in the literature for the pulse trains generated by the IFC. The simplest technique counts pulses in time bins to create a coarse time structure of the pulse train and apply standard algorithms on the vector space representation. Alvarado et al. [12] used this approach to solve the heartbeat classification problem with linear discriminant classifiers and binned pulses as features. McCormick [16] proposed asynchronous finite state machines

G. Nallathambi is with the Department of Electrical and Computer Engineering, Gainesville, Florida, 32611, USA (e-mail: gabriel_n,@ymail.com).

J. C. Principe is with the Department of Electrical and Computer Engineering, Gainesville, Florida, 32611, USA (e-mail: principe@cnel.ufl.edu).

The work of the authors is funded via the grant DARPA N66001-15-1-4054 and NSF EAGER 1723366.

to perform piecewise linear operations and reconstruct binary codes from input pulses. Signal processing is performed on the binary code followed by conversion back to pulses. Nallathambi and Principe [13] applied attribute grammars and automata directly to the pulse timing for performing non-numeric processing of pulse trains and identify QRS complexes in the electrocardiogram (ECG) signal with high accuracy. In the neuroscience literature, the pulse trains created by neurons are modeled as stochastic point processes [17], and many machine learning techniques are used to compute with pulses [18], [19]. While these works on pulse trains advanced signal representation and processing, there is a need for developing arithmetic operators for IFC pulse trains under a deterministic framework, i.e. assuming the signal is created from a deterministic source and the conversion is also deterministic, as used in sampling theory, which is the focus of this paper.

The main contributions of this paper are as follows. First, a theoretical framework for performing basic signal processing operations such as addition, multiplication, and convolution is derived. Secondly, algorithms for online implementation of pulse-based arithmetic and convolution is proposed. Together, these developments enable direct processing of pulse trains without signal reconstruction. Due to the sparse representation of the IFC sampler, the arithmetic operations have limited accuracy near the noise floor and low amplitude regions, but still effectively process the relevant information in the signal.

The ability of selectively capturing and processing the semantic information in the signal is important in many continuous and event monitoring applications for the Internet of Things (IoT) and mobile wireless sensor networks [20]. Applications where the goal is detection or classification of vital events and not necessarily signal reconstruction, are ideal for the proposed pulse-based algorithms, which represent the features of interest in the signal while suppressing the background noise.

The performance of the proposed algorithms is studied by quantifying the variations in instantaneous occurrence of pulses. The effect of IFC parameters and the efficiency of the approach in processing semantic regions of interest is demonstrated using synthetic data. Comparisons are performed with digital processing of reconstructed pulse trains. An application of noise reduction in ECG signal is demonstrated with sparse pulse representation while preserving the sematic features of interest. Matlab scripts for the key algorithms are made available in [21].

The rest of the paper is organized as follows: Section II describes the IFC in detail and presents the related works on pulse-based signal processing. Section III derives the theoretical framework for operating with pulse trains to perform addition, multiplication, and convolution. Section IV proposes algorithms for online implementation of the theoretical framework. Section V describes the datasets and performance metrics used for validation. Section VI quantifies the performance of the algorithms using synthetic and natural data. Section VII discusses the possibilities offered by the present work.

## II. INTEGRATE AND FIRE CONVERTER

Pulse based arithmetic and signal processing are developed with the objective of performing computation on analog signals using representations with digital amplitude but analog time. Pulse trains are waveforms where the information is contained in the timing of pulses instead of amplitude. The use of pulses for signal processing is not a new idea. Early efforts include works on arithmetic using pulse encoding methods such as pulse rate, width, edge, burst, phase, delay, and amplitude [22]–[25].

Since its inception, many studies such as pulse-based population encoding for single or multiple sensors in video processing [8], [9], [26], time-embedding based on the inter pulse intervals [27], learned input-output mappings based on a stochastic model for the events [18], [19], stochastic point process models [17], projections into reproducing kernel Hilbert spaces [28], and others [29] based on pulse streams have been proposed. Based on these works, various implementation schemes for pulse signal processing are proposed using magnetic cores [30], reconfigurable analog systems [31], fourth order palmo filter [32], etc. The trends in silicon technology with a decrease in voltage and an increase in speed are making pulse-based representations more appealing.

In this paper, we focus our discussion on the biphasic integrate and fire converter (IFC), which converts real world analog signals to analog time between pulses. The IFC output encodes information on both the timing of the pulses (analog) and polarity of pulses (digital). The methodology developed in this work can be easily applied to single polarity pulse trains as well.

Feichtinger et al. [10] among others [33]–[36] studied the use of IFC as a replacement for ADC and showed that the integrate and fire model can be used as a representation of the analog signal. Their work proved that the output of IFC, which codifies the variation of the integral of the signal, recovers the bandlimited analog signal with an error as small as required. One of the features of this approach versus Asynchronous Sigma Delta Modulation is that the achievable data rates are similar (or better) than the corresponding Nyquist samplers.

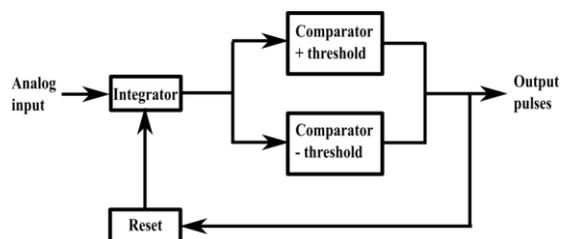

*Fig. 1. Block diagram of the biphasic integrate and fire analog to pulse converter.*

The IFC block diagram used in this paper is shown in Fig. 1. The analog input is integrated, and the result is compared against two thresholds. When either the positive or negative threshold $\theta$ is reached, a pulse is generated at time $t_k$ with positive or negative polarity $p_k$ respectively. Unlike the integrate and fire neuron model, two thresholds are used to reduce the pulse rate substantially [37]. Fundamentally, each pulse interval satisfies the condition

$$\theta = \int_{t_k}^{t_{k+1}} x(t) e^{-\alpha(t_{k+1}-t)} dt \quad (1)$$

where $\alpha$ is the rate of decay of the integrator, and $\theta$ is the threshold of the IFC. The pulse timings, the threshold and the rate of decay completely define the IFC pulse train output.

The IFC pulse train representation is rather different from discrete time representations. Pulses occur asynchronously in time, controlled by the amplitude of the analog signal, and the values of $\theta$ and $\alpha$. For this reason, the density of pulses is not a constant, with more pulses occurring in the large amplitude region of the analog signal, and fewer pulses appearing in the low amplitude portions of the analog signal. This creates a fundamental constraint for reconstruction and processing of pulse trains. Feichtinger et al. [15] studied the reconstruction of the analog signal from the pulses using frame theory and showed that it is possible to approximately reconstruct a bandlimited signal in $L^\infty$ norm with an error proportional to the threshold $\theta$. In [35] a simpler procedure employing finite bandlimited spaces is presented based on least squares using splines or Fourier bases such that $\hat{x}(t) = \sum_{k=1}^{M} a_k \phi_k(t)$, where $a_k$ is given by the linear regression $\vec{\theta} = S\vec{a}$, S is obtained by integrating the basis set over the reconstruction interval, and $\|x(t) - \hat{x}(t)\|_\infty \leq C\theta$ where $C$ is a constant solely dependent on the window of analysis and the choice of the bases functions. While it is possible to decrease the threshold $\theta$ to an arbitrarily small value, which reduces the reconstruction error, the pulse densities become well beyond what Nyquist rate requires. However, the present work focusses on applications where representation of semantic information content is important and the goal is not necessarily signal reconstruction but classification or interpretation of signal features in the pulse train.

We explain next a theoretical framework for performing basic signal processing operations such as arithmetic and convolution directly on pulse trains. Moreover, algorithms to implement these operators are also proposed, where the processing of information is online and entirely in the time domain as the inputs and output of the system are pulse trains.

### III. Theory of Pulse Signal Processing for IFC

IFC maps a continuous time, continuous amplitude signal into the structure of train of pulses in analog time such that the distance between any consecutive pulses $t_k$ and $t_{k+1}$ is fundamentally constrained by the threshold $\theta$, which controls the density of pulses; therefore, any arithmetic operation on pulse trains (addition or multiplication of pulses) also must be constrained by $\theta$. From eqn. 1, it is observed that $\theta$ is equal to the leaky area under $x(t)$ between $t_k$ and $t_{k+1}$ where the rate of decay is given by $\alpha$. Hence, any operation on pulse trains corresponds to equivalent operations on underlying areas.

Intuitively, this is straightforward to determine from eqn. 1, which is rewritten as $x(t) = \frac{\theta}{t_{k+1}-t_k}$ by assuming the rate of decay to be zero and $x(t)$ to be constant between $t_k$ and $t_{k+1}$. Hence, online addition of continuous time signals, $f(t) = c(t) + h(t)$ is expressed as $t_{f_{n+1}} - t_{f_n} = \frac{(t_{c_{n+1}} - t_{c_n})(t_{h_{n+1}} - t_{h_n})}{(t_{c_{n+1}} - t_{c_n}) + (t_{h_{n+1}} - t_{h_n})}$, where $t_{f_k}, t_{c_k}$, and $t_{h_k}$ are the $k^{th}$ pulse of $f(t)$, $c(t)$, and $h(t)$ respectively. This shows that mathematical operations on areas under the curve, which is related to amplitude of analog signal, are equivalent to operations on continuous time differences in consecutive pulses; alternatively, arithmetic operations in pulse time differences when constrained by $\theta$ correspond to equivalent operations on areas under the curve of analog signals. Therefore, $A_1(t_j, t_{j+1}) \phi A_2(t_j, t_{j+1}) = \theta$, where $\phi$ represents the operator (+ or *) on pulse trains, $t_j$ is the resulting pulses due to the operation, $A_i(t_j, t_{j+1})$ is the underlying area of the $i^{th}$ pulse train ($i = 1, 2$) during $(t_j, t_{j+1})$;

Based on this intuition, theoretical framework is proposed to perform arithmetic and convolution using IFC pulse trains. An overlapping time interval in the operands $(t_a, t_b)$ is used within the pulse intervals for deriving the theorems and generalizing the results. The choice of values for $t_a$ and $t_b$ are discussed in the next section.

The framework solves the equation $A_1(t_a, t_b) \phi A_2(t_a, t_b) = \mu\theta$ for $\mu$ and $t_j$, where $\mu$ is the area in terms of $\theta$ resulting from the operation and the pulse timing $t_j$ occurs when $\mu = 1$. To solve deterministically the above equation, it is assumed that the input signal is constant between $t_k$ and $t_{k+1}$. This simple signal model is shown to be sufficient in preserving the semantic information in the signal and suitable for many IoT applications that do not require accurate signal reconstruction. For completeness, the corresponding error bounds due to this assumption are presented. In this section, theoretical framework is derived for calculating an instance of $t_j$ with no carryovers, and we propose algorithms for updating area recursively with carryovers in section IV.

*Observation 1:* If $\theta = \int_{t_k}^{t_{k+1}} x(t) e^{-\alpha(t_{k+1}-t)} dt$ and $x(t)$ is constant between $t_k$ and $t_{k+1}$, then $\xi\theta = \int_{t_a}^{t_b} x(t) e^{-\alpha(t_{k+1}-t)} dt$, where $\xi = \frac{g(t_{k+1}-t_a) - g(t_{k+1}-t_b)}{g(t_{k+1}-t_k)}$, $t_k \leq t_a < t_b \leq t_{k+1}$, and $g(t) = 1 - e^{-\alpha t}$. This observation is critical in the derivation of theorems for arithmetic as it enables generalization of the results in any interval $(t_a, t_b)$ between two consecutive pulses.

From the mean value theorem [38], it can be shown that the error $\delta_x$ due to the assumption of $x(t)$ being a constant $c_x$ between $t_k$ and $t_{k+1}$ is bounded by $\frac{\min_{z \in [t_a, t_b]} |x(z) - c_x|}{e^{\alpha(t_{k_{n+1}} - t_a)}} \leq \frac{|\delta_x|}{(t_b - t_a)} \leq \frac{\max_{z \in [t_a, t_b]} |x(z) - c_x|}{e^{\alpha(t_{k_{n+1}} - t_b)}}$. Hence, the error depends on the length of the interval $(t_b - t_a)$ and the deviation of $x(t)$ from $c_x$ in $[t_a, t_b]$.

### A. Theorem 1: Online addition of pulse trains

Consider two continuous time, continuous amplitude signals $x(t)$ and $y(t)$ corresponding to augend and addend pulse trains respectively. Suppose the augend pulses occur at $t_{x_j}$ with polarity $p_{x_j}$, addend pulses occur at $t_{y_j}$ with polarity $p_{y_j}$, and the new pulses due to their sum occur at $t_{s_j}$ with polarity $p_{s_j}$ such that $t_{x_n}, t_{y_n} \leq t_a < t_b \leq t_{x_{n+1}}, t_{y_{n+1}}$, then it is shown that $t_{s_{n+1}} = \frac{-1}{\alpha} \ln\{1 - KS\} + t_{s_n}$ and $p_{s_{n+1}} = sgn(\eta)$ assuming

$x(t)$ and $y(t)$ to be constant between consecutive pulses, where $sgn(x)$ is the signum function,

$$K = \frac{sgn(\eta)p_{x_{n+1}}}{p_{y_{n+1}}},$$

$$S = \frac{g(t_{x_{n+1}}-t_{x_n})g(t_{y_{n+1}}-t_{y_n})}{p_{x_{n+1}}g(t_{x_{n+1}}-t_{x_n})+p_{x_{n+1}}g(t_{y_{n+1}}-t_{y_n})},$$

$$\eta = \frac{p_{x_{n+1}}[g(t_{x_{n+1}}-t_a)-g(t_{x_{n+1}}-t_b)]}{g(t_{x_{n+1}}-t_{x_n})} + \frac{p_{y_{n+1}}[g(t_{y_{n+1}}-t_a)-g(t_{y_{n+1}}-t_b)]}{g(t_{y_{n+1}}-t_{y_n})}.$$

*Proof:*

Addition of pulse trains corresponds to the sum of the underlying areas. Based on the proposed framework, the addition operation on pulse trains is performed by solving eqn. 2 for $\mu$ and $t_{s_{n+1}}$. Assuming $x(t)$ and $y(t)$ to be constant between consecutive pulses, we have

$$\int_{t_a}^{t_b} xe^{-\alpha(t_{x_{n+1}}-t)}dt + \int_{t_a}^{t_b} ye^{-\alpha(t_{y_{n+1}}-t)}dt \quad (2)$$
$$= \mu \int_{t_{s_n}}^{t_{s_{n+1}}} [x+y]e^{-\alpha(t_{s_{n+1}}-t)}dt$$

Using Observation 1, eqn. 2 is written as $p_{x_{n+1}}u\theta + p_{y_{n+1}}d\theta = \eta\theta$, where $u = \frac{g(t_{x_{n+1}}-t_a)-g(t_{x_{n+1}}-t_b)}{g(t_{x_{n+1}}-t_{x_n})}$, $d = \frac{g(t_{y_{n+1}}-t_a)-g(t_{y_{n+1}}-t_b)}{g(t_{y_{n+1}}-t_{y_n})}$, $\mu = |\eta|$, and $p_{s_{n+1}} = sgn(\eta)$. Moreover, $y$ is expressed in terms of $x$ as $y = \frac{p_{x_{n+1}}xg(t_{x_{n+1}}-t_{x_n})}{p_{y_{n+1}}g(t_{y_{n+1}}-t_{y_n})}$ since $p_{x_{n+1}} \int_{t_{x_n}}^{t_{x_{n+1}}} xe^{-\alpha(t_{x_{n+1}}-t)}dt = p_{y_{n+1}} \int_{t_{y_n}}^{t_{y_{n+1}}} ye^{-\alpha(t_{y_{n+1}}-t)}dt$.

By substituting $y$ and $\mu$ in eqn. 2, we obtain

$$g(t_{s_{n+1}} - t_{s_n}) = \frac{Kg(t_{x_{n+1}}-t_{x_n})g(t_{y_{n+1}}-t_{y_n})}{p_{x_{n+1}}g(t_{x_{n+1}}-t_{x_n}) + p_{y_{n+1}}g(t_{y_{n+1}}-t_{y_n})} \quad (3)$$

where $K = \frac{sgn(\eta)p_{x_{n+1}}}{p_{y_{n+1}}}$, $K \in (1, -1)$.

Thus, from eqn. 3, the polarity and timing of the sum of the pulse trains is given by $p_{s_{n+1}} = sgn(\eta)$ and $t_{s_{n+1}} - t_{s_n} = \frac{-1}{\alpha}\ln\left\{1 - \frac{Kg(t_{x_{n+1}}-t_{x_n})g(t_{y_{n+1}}-t_{y_n})}{p_{x_{n+1}}g(t_{x_{n+1}}-t_{x_n})+p_{x_{n+1}}g(t_{y_{n+1}}-t_{y_n})}\right\}$ respectively.

The error $\delta_{x+y}$ resulting from the addition of two pulse trains, with the assumption of $x(t) = c_x$ between consecutive pulses, is given by the quadrature formula [39], $\delta_{x+y} = \{(\delta_x)^2 + (\delta_y)^2\}^{\frac{1}{2}}$ and bounded by $\frac{\min_{z\in[t_a,t_b]}(x(z)-c_x)^2}{e^{2\alpha(t_{x_{n+1}}-t_a)}} + \frac{\min_{z\in[t_a,t_b]}(y(z)-c_y)^2}{e^{2\alpha(t_{y_{n+1}}-t_a)}} \leq \frac{\delta_{x+y}^2}{(t_b-t_a)^2} \leq \frac{\max_{z\in[t_a,t_b]}(x(z)-c_x)^2}{e^{2\alpha(t_{x_{n+1}}-t_b)}} + \frac{\max_{z\in[t_a,t_b]}(y(z)-c_y)^2}{e^{2\alpha(t_{y_{n+1}}-t_b)}}$.

### B. Theorem 2: Online multiplication of pulse trains

Consider two continuous time, continuous amplitude signals $x(t)$ and $y(t)$ corresponding to multiplicand and multiplier pulse trains respectively and let $r(t) = 1$ correspond to the identity (reference) pulse train. Suppose the multiplicand pulses occur at $t_{x_j}$ with polarity $p_{x_j}$, multiplier pulses occur at $t_{y_j}$ with polarity $p_{y_j}$, reference pulses occur at $t_{r_j}$ and the product of the two pulse trains occur at $t_{p_j}$ with polarity $p_{p_j}$ such that $t_{x_n}, t_{y_n} \leq t_a < t_b \leq t_{x_{n+1}}, t_{y_{n+1}}$ then it is shown that $t_{p_{n+1}} = \frac{-1}{\alpha}\ln\{1 - P\} + t_{p_n}$ and $p_{p_{n+1}} = sgn(\eta)$ assuming $x(t)$ and $y(t)$ to be constant between consecutive pulses, where

$$P = \frac{g(t_{x_{n+1}}-t_{x_n})g(t_{y_{n+1}}-t_{y_n})}{g(t_{r_{n+1}}-t_{r_n})}, \text{ and}$$

$$\eta = \left[\frac{p_{x_{n+1}}p_{y_{n+1}}g(t_{r_{n+1}}-t_{r_n})}{g(t_{r_{n+1}}-t_a)-g(t_{r_{n+1}}-t_b)}\right] \cdot \left[\frac{g(t_{x_{n+1}}-t_a)-g(t_{x_{n+1}}-t_b)}{g(t_{x_{n+1}}-t_{x_n})}\right].$$

$$\left[\frac{g(t_{y_{n+1}}-t_a)-g(t_{y_{n+1}}-t_b)}{g(t_{y_{n+1}}-t_{y_n})}\right].$$

*Proof.*

Multiplication of pulse trains requires normalization by the identity pulse train, which is a periodic pulse train corresponding to $r(t) = 1$. Hence, by assuming $x(t)$ and $y(t)$ to be constant between consecutive pulses, we have

$$\frac{\int_{t_a}^{t_b} xe^{-\alpha(t_{x_{n+1}}-t)}dt \int_{t_a}^{t_b} ye^{-\alpha(t_{y_{n+1}}-t)}dt}{\int_{t_a}^{t_b} e^{-\alpha(t_{r_{n+1}}-t)}dt} \quad (4)$$
$$= \mu \int_{t_{p_n}}^{t_{p_{n+1}}} [xy]e^{-\alpha(t_{p_{n+1}}-t)}dt$$

Using Observation 1, eqn. 4 is written as $\frac{(p_{x_{n+1}}u\theta)(p_{y_{n+1}}d\theta)}{p_{r_{n+1}}r\theta} = \eta\theta$, where $u = \frac{g(t_{x_{n+1}}-t_a)-g(t_{x_{n+1}}-t_b)}{g(t_{x_{n+1}}-t_{x_n})}$, $d = \frac{g(t_{y_{n+1}}-t_a)-g(t_{y_{n+1}}-t_b)}{g(t_{y_{n+1}}-t_{y_n})}$, $r = \frac{g(t_{r_{n+1}}-t_a)-g(t_{r_{n+1}}-t_b)}{g(t_{r_{n+1}}-t_{r_n})}$, $\mu = |\eta|$ and $p_{p_{n+1}} = sgn(\eta)$.

Substituting $\mu$ in eqn. 4, we obtain

$$g(t_{p_{n+1}} - t_{p_n}) = \frac{g(t_{x_{n+1}}-t_{x_n})g(t_{y_{n+1}}-t_{y_n})}{g(t_{r_{n+1}}-t_{r_n})} \quad (5)$$

Thus, from eqn. 5, the polarity and timing of the product of the pulse trains is given by $p_{p_{n+1}} = sgn(\eta)$ and $t_{p_{n+1}} - t_{p_n} = \frac{-1}{\alpha}\ln\left\{1 - \frac{g(t_{x_{n+1}}-t_{x_n})g(t_{y_{n+1}}-t_{y_n})}{g(t_{r_{n+1}}-t_{r_n})}\right\}$ respectively.

The error $\delta_{xy}$ due to the product of two pulse trains, with the assumption of $x(t) = c_x$ between consecutive pulses, is given by the formula [39], $\delta_{xy} = \left|\frac{K_1K_2\theta}{K_3}\right|\left[\left(\frac{\delta_x}{K_1\theta}\right)^2 + \left(\frac{\delta_y}{K_2\theta}\right)^2\right]^{\frac{1}{2}}$ where $\int_{t_a}^{t_b} x(t)e^{-\alpha(t_{x_{n+1}}-t)}dt = K_1\theta$, $\int_{t_a}^{t_b} y(t)e^{-\alpha(t_{y_{n+1}}-t)}dt = K_2\theta$, $\int_{t_a}^{t_b} e^{-\alpha(t_{r_{n+1}}-t)}dt = K_3\theta$, and bounded by $\frac{K_2^2 \min_{z\in[t_a,t_b]}(x(z)-c_x)^2}{e^{2\alpha(t_{x_{n+1}}-t_a)}} + \frac{K_1^2 \min_{z\in[t_a,t_b]}(y(z)-c_y)^2}{e^{2\alpha(t_{y_{n+1}}-t_a)}} \leq \left(\frac{K_3}{t_b-t_a}\right)^2 \delta_{xy}^2 \leq \frac{K_2^2 \max_{z\in[t_a,t_b]}(x(z)-c_x)^2}{e^{2\alpha(t_{x_{n+1}}-t_b)}} + \frac{K_1^2 \max_{z\in[t_a,t_b]}(y(z)-c_y)^2}{e^{2\alpha(t_{y_{n+1}}-t_b)}}$.

### C. Theorem 3: Online convolution of pulse trains

Consider two continuous time, continuous amplitude signals $x(t)$ and $y(t)$ corresponding to pulse train X and pulse train Y respectively and let $r(t) = 1$ correspond to the identity (reference) pulse train. Suppose the pulses of X occur at $t_{x_j}$ with polarity $p_{x_j}$, pulses of Y occur at $t_{y_j}$ with polarity $p_{d_j}$, reference pulses occur at $t_{r_j}$ and the convolution of the two pulse trains occur at $t_{c_j}$ with polarity $p_{c_j}$ such that $t_{x_n}, t_{y_n} \leq t_a < t_b \leq t_{x_{n+1}}, t_{y_{n+1}}$ then it is shown that $t_{c_{n+1}} = \frac{-1}{\alpha}\ln\{1 - P\} + t_{c_n}$ and

$p_{c_{n+1}} = sgn(\eta)$ assuming $x(t)$ and $y(t)$ to be constant between consecutive pulses, where $P = \frac{g(t_{x_{n+1}}-t_{x_n})g(t_{y_{n+1}}-t_{y_n})}{g(t_{r_{n+1}}-t_{r_n})(T_2-T_1)}$, $T = T_2 - T_1$ is the period of intersection of the two pulse trains,

$$\eta = \frac{-1}{\alpha}\left[\frac{p_{x_{n+1}}p_{y_{n+1}}g(t_{r_{n+1}}-t_{r_n})}{g(t_{r_{n+1}}-t_a)-g(t_{r_{n+1}}-t_b)}\right]\cdot\left[\frac{g(t_{x_{n+1}}-t_a)-g(t_{x_{n+1}}-t_b)}{g(t_{x_{n+1}}-t_{x_n})}\right]\cdot\left[\frac{g(t_{y_{n+1}}-t_a+\lambda_2)-g(t_{y_{n+1}}-t_b+\lambda_2)-g(t_{y_{n+1}}-t_a+\lambda_1)+g(t_{y_{n+1}}-t_b+\lambda_1)}{g(t_{y_{n+1}}-t_{y_n})}\right],$$

and $\lambda = \lambda_2 - \lambda_1$ is the time offset (shift).

*Proof.*

Convolution of pulse trains requires convolution of underlying areas and hence, we have

$$\frac{\int_{t_a}^{t_b} x e^{-\alpha(t_{x_{n+1}}-t)}dt \otimes \int_{t_a}^{t_b} y e^{-\alpha(t_{y_{n+1}}-t)}dt}{\int_{t_a}^{t_b} e^{-\alpha(t_{r_{n+1}}-t)}dt} \quad (6)$$

$$= \mu \int_{t_{c_n}}^{t_{c_{n+1}}} [x \otimes y] e^{-\alpha(t_{c_{n+1}}-t)}dt$$

Eqn. 6 is expressed as follows:

$$\frac{\int_{t_a}^{t_b} e^{-\alpha(t_{x_{n+1}}-\tau)}d\tau}{\int_{t_a}^{t_b} e^{-\alpha(t_{r_{n+1}}-\tau)}d\tau}\left[\frac{1}{\alpha}\int_{\lambda=\lambda_1}^{\lambda_2}\left(e^{-\alpha(t_{y_{n+1}}-t_b+\lambda)}\right.\right.$$
$$\left.\left.- e^{-\alpha(t_{y_{n+1}}-t_b+\lambda)}\right)d\lambda\right] \quad (7)$$

$$= \mu \int_{t_{c_n}}^{t_{c_{n+1}}}\left[\int_{\beta=T_1}^{T_2} d\beta\right]e^{-\alpha(t_{c_{n+1}}-t)}dt$$

Using Observation 1, the above equation is written as $\frac{(p_{x_{n+1}}u\theta)(p_{y_{n+1}}d\theta)}{\alpha p_{r_{n+1}}r\theta} = \eta\theta$, where $u = \frac{g(t_{x_{n+1}}-t_a)-g(t_{x_{n+1}}-t_b)}{g(t_{x_{n+1}}-t_{x_n})}$, $r = \frac{g(t_{r_{n+1}}-t_a)-g(t_{r_{n+1}}-t_b)}{g(t_{r_{n+1}}-t_{r_n})}$, $\mu = |\eta|$, $p_{c_{n+1}} = sgn(\eta)$, $d = \frac{g(t_{y_{n+1}}-t_a+\lambda_2)-g(t_{y_{n+1}}-t_b+\lambda_2)-g(t_{y_{n+1}}-t_a+\lambda_1)+g(t_{y_{n+1}}-t_b+\lambda_1)}{g(t_{y_{n+1}}-t_{y_n})}$.

By substituting $\mu$ in eqn. 7, we obtain

$$g(t_{c_{n+1}} - t_{c_n}) = \frac{g(t_{x_{n+1}} - t_{x_n})g(t_{y_{n+1}} - t_{y_n})}{(T_2 - T_1)g(t_{r_{n+1}} - t_{r_n})} \quad (8)$$

Thus, from eqn. 8, the polarity and timing of the product of the pulse trains is given by $p_{c_{n+1}} = sgn(\eta)$ and $t_{c_{n+1}} - t_{c_n} = \frac{-1}{\alpha}\ln\left\{1 - \frac{g(t_{x_{n+1}}-t_{x_n})g(t_{y_{n+1}}-t_{y_n})}{(T_2-T_1)g(t_{r_{n+1}}-t_{r_n})}\right\}$ respectively.

The error $\delta_{x\otimes y}$ due to the convolution of two pulse trains is bounded by 
$$\frac{K_2^2 \min_{z\in[t_a,t_b]}(x(z)-c_x)^2}{e^{2\alpha(t_{x_{n+1}}-t_a)}} + \frac{K_1^2 \min_{z'\in[\lambda_1-t_b,\lambda_2-t_a]}(y(z')-c_y)^2}{e^{2\alpha(t_{y_{n+1}}-(\lambda_1-t_b))}} \leq \left(\frac{K_3}{(t_b-t_a)(\lambda_2-\lambda_1)}\right)^2 \delta_{x\otimes y}^2 \leq \frac{K_2^2 \max_{z\in[t_a,t_b]}(x(z)-c_x)^2}{e^{2\alpha(t_{x_{n+1}}-t_b)}} + \frac{K_1^2 \max_{z'\in[\lambda_1-t_b,\lambda_2-t_a]}(y(z')-c_y)^2}{e^{2\alpha(t_{y_{n+1}}-(\lambda_2-t_a))}}.$$

This framework computes one instance of output pulse $t_j$ due to an operation. In the next section, online algorithms for pulse trains extend the theoretical framework, where the focus is on selecting $t_a$ and $t_b$, and on the recursive computation of areas.

## IV. Algorithms for Pulse Train Arithmetic

### A. Pulse arithmetic algorithm

In Table I, the algorithm for computing arithmetic of two pulse trains is presented. The time interval $(t_a, t_b)$ may be selected to shift forward in fixed or variable windows within a pulse interval. Sliding $t_a$ and $t_b$ in fixed intervals requires prior knowledge of the minimum value of the inter-pulse interval to ensure the observation window lies within $(t_k, t_{k+1})$ as per Observation 1. In this paper, to ensure online implementation, variable windows are used where the interval shifts forward at the arrival of every new pulse in the operands. For instance, let pulse train X have pulses at {1s, 3s} and pulse train Y have pulses at {2s, 3s}, then the overlapping time interval $(t_a, t_b)$ is selected in the following order for calculation of areas: (0, 1s), (1s, 2s), and (2s, 3s).

TABLE I
ALGORITHM FOR PULSE TRAIN ARITHMETIC

1. Select computation time points: $t_a, t_b$
2. Obtain pulse intervals corresponding to computation time points: $t_{x_{n+1}}, t_{x_n}, t_{y_{n+1}}, t_{y_n}$
3. Calculate $\eta$. $\left\{\text{Addition: } \eta = \frac{p_{x_{n+1}}[g(t_{x_{n+1}}-t_a)-g(t_{x_{n+1}}-t_b)]}{g(t_{x_{n+1}}-t_{x_n})} + \frac{p_{y_{n+1}}[g(t_{y_{n+1}}-t_a)-g(t_{y_{n+1}}-t_b)]}{g(t_{y_{n+1}}-t_{y_n})}\right\};$

$$\left\{\text{Multiplication: } \eta = \left[\frac{p_{x_{n+1}}p_{y_{n+1}}g(t_{r_{n+1}}-t_{r_n})}{g(t_{r_{n+1}}-t_a)-g(t_{r_{n+1}}-t_b)}\right]\cdot\left[\frac{g(t_{x_{n+1}}-t_a)-g(t_{x_{n+1}}-t_b)}{g(t_{x_{n+1}}-t_{x_n})}\right]\cdot\left[\frac{g(t_{y_{n+1}}-t_a)-g(t_{y_{n+1}}-t_b)}{g(t_{y_{n+1}}-t_{y_n})}\right]\right\}.$$

4. Compute output pulse timing and polarity: $t_k, p_k$

Calculate $g(t_{p_{n+1}} - t_{p_n})$. $\left\{\text{Addition: } g(t_{p_{n+1}} - t_{p_n}) = \frac{Kg(t_{x_{n+1}}-t_{x_n})g(t_{y_{n+1}}-t_{y_n})}{p_{x_{n+1}}g(t_{x_{n+1}}-t_{x_n})+p_{y_{n+1}}g(t_{y_{n+1}}-t_{y_n})}\right\};$
$\left\{\text{Multiplication: } g(t_{p_{n+1}} - t_{p_n}) = \frac{g(t_{x_{n+1}}-t_{x_n})g(t_{y_{n+1}}-t_{y_n})}{g(t_{r_{n+1}}-t_{r_n})}\right\}.$

**while** $|\eta + \eta_{ex}| \geq 1$
$\quad t_k = \frac{-1}{\alpha}\ln\{1 - g(t_{p_{n+1}} - t_{p_n})|sgn(\eta) - \eta_{ex}|\} + t_{k-1} + t_{ex}$
$\quad p_k = sgn(\eta)$
$\quad$ *Update*:
$\quad t_a = t_k$
$\quad$ *Calculate* $\eta$
$\quad \eta_{ex}, t_{ex} \to 0, \; k \to k+1$
**end**

$\eta_{ex} \to \eta + \eta_{ex}$
$t_{ex} \to t_b - t_k$

To recursively update the area, carryovers in both timing and area namely excess time $(t_{ex})$ and excess area $(\eta_{ex})$ are introduced. To perform arithmetic and convolution, excess time and excess area are included in the calculation of total area and new pulse timing at every computation as shown in Table I. The carryovers ensure the output pulse timings always fall within the current observation window $(t_a, t_b)$.

Thus the process involves selection of observation window and its associated pulse intervals, and calculation of the total area $\eta + \eta_{ex}$ as per Table I. When the total area exceeds +1 or -1, the pulse occurs at that time instant $t_k$ with corresponding polarity $p_k$.

## B. Approximations

As the values of inter-pulse intervals are inversely proportional to signal amplitude with high inter-pulse intervals corresponding to noise floor, $g(m)$ is approximated by $\alpha m$ without degrading performance. Then, the arithmetic equations in Table I for $\eta$ and $t_k$ is simplified as shown in Table II. Unlike Table I, the simplified equations for $\eta$ and $t_k$ rely directly on the inter-pulse intervals; therefore, the real-time hardware implementation of pulse-based systems is straightforward.

TABLE II
SIMPLIFIED ALGORITHM FOR PULSE TRAIN ARITHMETIC

1. Select computation time points: $t_a, t_b$
2. Obtain pulse intervals corresponding to computation time points: $t_{x_{n+1}}, t_{x_n}, t_{y_{n+1}}, t_{y_n}$
3. Calculate $\eta$.
$$\left\{\text{Addition}: \eta = \frac{p_{x_{n+1}}(t_b - t_a)}{t_{x_{n+1}} - t_{x_n}} + \frac{p_{y_{n+1}}(t_b - t_a)}{t_{y_{n+1}} - t_{y_n}}\right\};$$
$$\left\{\text{Multiplication}: \eta = \left[\frac{p_{x_{n+1}} p_{y_{n+1}}(t_b - t_a)(t_{r_{n+1}} - t_{r_n})}{(t_{x_{n+1}} - t_{x_n})(t_{y_{n+1}} - t_{y_n})}\right]\right\}$$
4. Compute output pulse timing and polarity: $t_k, p_k$

Calculate $t_{p_{n+1}} - t_{p_n} = \frac{t_b - t_a}{|\eta|}$.

while $|\eta + \eta_{ex}| \geq 1$
$t_k = (t_{p_{n+1}} - t_{p_n})|sgn(\eta) - \eta_{ex}| + t_{k-1} + t_{ex}$
$p_k = sgn(\eta)$
Update:
$\eta \rightarrow \eta - (sgn(\eta) - \eta_{ex})$
$\eta_{ex}, t_{ex} \rightarrow 0, \; k \rightarrow k + 1$
end

$\eta_{ex} \rightarrow \eta + \eta_{ex}$
$t_{ex} \rightarrow t_b - t_k$

## C. Pulse convolution algorithm

The algorithm for computing the convolution of two pulse trains X and Y is presented in Table III. The pulse timings of Y are reversed and shifted by $\lambda$, which is given by the minimum distance required for pulses of Y to reach one of the pulses of X upon shifting. Overlap between the two pulse trains after shifting is computed. Unlike pulse arithmetic, there is a vector of computation time points $t_{a_i}, t_{b_i}$ and an associated vector of pulse intervals corresponding to all pulses in the region of overlap. Each element of these vectors results in an area $\eta_i$ and the total area resulting during a shift operation is given by the sum of $\eta_i$'s. When the total area exceeds +1 or -1, the component of $\eta$ at which this occurs determines the timing $t_k$ and polarity $p_k$ of the pulse resulting from the convolution of pulse trains. Similar to Table II, the equations for convolution can be approximated by $\eta = \left[\frac{p_{x_{n+1}} p_{y_{n+1}}(t_b - t_a)(t_{r_{n+1}} - t_{r_n})(\lambda_2 - \lambda_1)}{(t_{x_{n+1}} - t_{x_n})(t_{y_{n+1}} - t_{y_n})}\right]$ and $t_{p_{n+1}} - t_{p_n} = \frac{(t_{x_{n+1}} - t_{x_n})(t_{y_{n+1}} - t_{y_n})}{(T_2 - T_1)(t_{r_{n+1}} - t_{r_n})}$.

## V. PERFORMANCE VALIDATION

### A. Performance measures

Measures based on inter pulse intervals: The quality of performance is evaluated in terms of measures of accuracy between instantaneous amplitude values $\hat{z}$ and $z$ calculated from the inter pulse intervals of the algorithmic pulse train output $\hat{Z}$ and desired pulse train output $Z$ respectively. The instantaneous amplitude $\hat{z}$ of a pulse train with consecutive pulses $t_k$ and $t_{k+1}$ is given by $\hat{z} = \frac{\theta}{t_{k+1} - t_k}$ and $\hat{z} = \frac{\theta \alpha}{1 - e^{-\alpha(t_{k+1} - t_k)}}$ for IFC with zero and non-zero rate of decay respectively. Peak signal to noise ratio ($PSNR$) and correlation coefficient ($r$) are the measures used to quantify the accuracy between instantaneous amplitude values of the algorithmic and desired outputs, where $PSNR = -10 log\left(\frac{\sum_{i=0}^{N}(\hat{z}_i - z_i)^2}{N(\hat{z}_{max} - \hat{z}_{min})^2}\right)$, $r = \frac{\sum_{i=1}^{N}(\hat{z}_i - m_{\hat{z}})(z_i - m_z)}{\sqrt{\sum_{i=1}^{N}(\hat{z}_i - m_{\hat{z}})^2}\sqrt{\sum_{i=1}^{N}(z_i - m_z)^2}}$, and $m_{\hat{z}}$ and $m_z$ are the sample mean of $\hat{z}$ and $z$ respectively.

TABLE III
ALGORITHM FOR PULSE TRAIN CONVOLUTION

1. Choose time offset $\lambda = \lambda_2 - \lambda_1$.
2. Shift the pulse timings of the reversed pulse train by $\lambda$.
3. Find the overlap $T = T_2 - T_1$ between the pulse trains after shifting.
4. Select computation time points $t_{a_i}, t_{b_i}$ for all pulses in the region of intersection.
5. Obtain pulse intervals corresponding to all computation time points in the region of intersection: $t_{x_{n+1_i}}, t_{x_{n_i}}, t_{y_{n+1_i}}, t_{y_{n_i}}$
6. Calculate $\eta_i$ for all computation time points $t_{a_i}, t_{b_i}$

$$\text{Convolution}: \eta_i = \frac{-1}{\alpha}\left[\frac{p_{x_{n+1_i}} p_{y_{n+1_i}} g(t_{r_{n+1_i}} - t_{r_{n_i}})}{g(t_{r_{n+1_i}} - t_{a_i}) - g(t_{r_{n+1_i}} - t_{b_i})}\right].$$

$$\left[\frac{g(t_{x_{n+1_i}} - t_{a_i}) - g(t_{x_{n+1_i}} - t_{b_i})}{g(t_{x_{n+1_i}} - t_{x_{n_i}})}\right].$$

$$\left[\frac{g(t_{y_{n+1_i}} - t_{a_i} + \lambda_2) - g(t_{y_{n+1_i}} - t_{b_i} + \lambda_2) - g(t_{y_{n+1_i}} - t_{a_i} + \lambda_1) + g(t_{y_{n+1_i}} - t_{b_i} + \lambda_1)}{g(t_{y_{n+1}} - t_{y_n})}\right]$$

7. Compute output pulse timing and polarity: $t_k, p_k$

Calculate $t_{c_i} = \frac{-1}{\alpha} \ln\left\{1 - \frac{g(t_{x_{n+1_i}} - t_{x_{n_i}})g(t_{y_{n+1_i}} - t_{y_{n_i}})}{(T_2 - T_1)g(t_{r_{n+1_i}} - t_{r_{n_i}})}\right\}$ for all computation time points.
Calculate $\eta = sum(\eta_1, \eta_2, \cdots, \eta_n)$

while $|\eta + \eta_{ex}| \geq 1$
Find first $j$ at which $|\eta_{ex} + sum(\eta_1, \eta_2, \cdots, \eta_j)| \geq 1$
$\sigma = sgn(\eta_j) - \eta_{ex} - sum(\eta_1, \eta_2, \cdots, \eta_{j-1})$
$t_k = \{|\eta_1 t_{c_1}| + \cdots + |\eta_{j-1} t_{c_{j-1}}| + |\sigma t_{c_2}|\} + t_{k-1} + t_{ex}$
$p_k = sgn(\sigma)$
Update:
$\eta_j \rightarrow \eta_j - \sigma$
$\eta_1, \eta_2, \cdots, \eta_{j-1} \rightarrow 0$
$\eta = sum(\eta_j, \eta_{j+1}, \cdots, \eta_n)$
$\eta_{ex}, t_{ex} \rightarrow 0, \; k \rightarrow k + 1$
end

$\eta_{ex} \rightarrow \eta + \eta_{ex}$
$t_{ex} \rightarrow \lambda_2 - t_k$

Region of analysis: As the pulse representation is dependent on the structure of the input, the analysis window is subdivided into four regions namely A, B, C, and D based on amplitude quartiles. Unlike conventional digital signal processing, pulse-based computation has relatively lower incidence of pulses near the noise floor (region A) and high pulse density in the other regions of interest. Hence it is necessary to quantify the performance in the individual regions and the overall performance is reported in terms of mean ± standard deviation of all regions.

Comparative studies: The focus of the paper is on processing the semantic information directly with pulse trains without signal reconstruction. However, to ensure completeness, the algorithmic pulse train output $\hat{Z}$ is reconstructed to get $\hat{z}_r(n)$, and compared with $z_r(n)$ obtained using digital arithmetic of $x_r(n)$ and $y_r(n)$, which are reconstructed from input pulse trains $X$ and $Y$ respectively as per [15]. The performance is assessed by computing $PSNR$ and $r$ between $\hat{z}_r(n)$ and $z_r(n)$.

### B. Data analysis

Synthetic data: Aperiodic pulse trains $X$ and $Y$ generated from 1V, 1Hz sinusoidal signals $x(t)$ and $y(t)$ respectively are selected to demonstrate the performance of the algorithm. Performance of the algorithm is quantified for variations in the IFC parameters. Two-sample t-test at 5% significance level is used to study the significant differences in the mean PSNR and pulse rate of the algorithms with and without approximations. Comparative studies are performed using the synthetic data in terms of the aforementioned performance measures.

Real data: An application of subtraction of baseline wander from ECG signal is used to demonstrate the feasibility of the proposed pulse-based algorithm for semantic information processing in continuous patient monitoring systems. Input ECG pulse train X is obtained by corrupting an ECG signal of 30-minute duration from MIT-BIH database [40] (dataset 100) with 300μV, 0.2Hz sinusoidal baseline wander, and converting to pulses with the IFC parameters chosen as in [13]. Pulse subtraction of pulse train Y corresponding to sinusoidal baseline wander from pulse train X is used to illustrate the semantic information processing and representation of relevant features of interest in ECG.

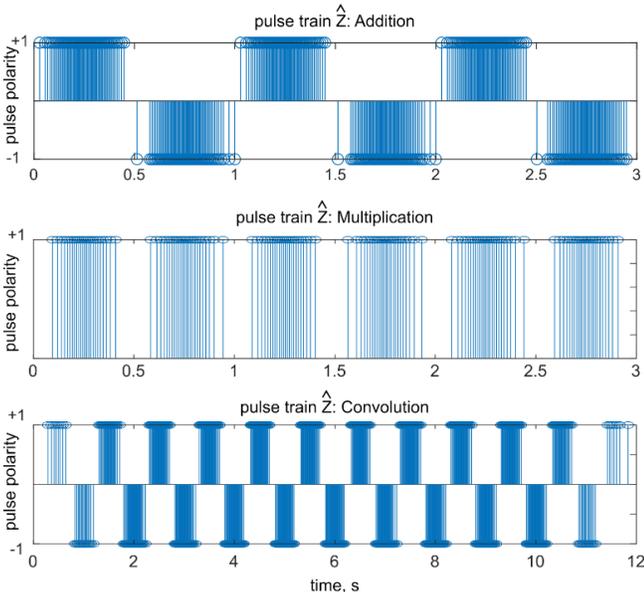

Fig. 2. Illustration of outputs of proposed algorithms. Convolution is shown for pulse trains X and Y of durations 10s and 2s respectively.

## VI. RESULTS

### A. Algorithmic performance in regions of analysis

The algorithmic outputs of synthetic data for pulse train addition, multiplication, and convolution with $\theta = 0.01$ and zero rate of decay is illustrated in Fig. 2. Performance analysis of pulse train addition shows PSNR, $r$, and pulse rate of 32.58 ± 18.35 dB, 0.92 ± 0.14, and 31.0 ± 36.57 pulses per second (p/s) respectively. The performance of algorithms is dependent on the regions of activity as demonstrated in Table IV for pulse train addition where PSNR is greater than 40dB with $r = 1$ at regions C and D, and less than 25dB with reduced pulse rate at regions A and B.

TABLE IV
PERFORMANCE IN THE REGIONS OF ANALYSIS

| Region | PSNR, dB | r | APR, p/s |
|---|---|---|---|
| Region A | 14.10 | 0.71 | 3.33 |
| Region B | 20.60 | 0.97 | 10.67 |
| Region C | 42.36 | 1.0 | 26 |
| Region D | 53.47 | 1.0 | 84 |

Likewise, pulse train multiplication has PSNR, $r$, and pulse rate of 31.45 ± 13.7 dB, 0.95 ± 0.07, and 12.25 ± 12.75 p/s respectively, while pulse train convolution has PSNR, $r$, and pulse rate of 29.61 ± 4.73 dB, 0.98 ± 0.03, and 12.38 ± 11.49 p/s respectively. These results demonstrate that the proposed algorithms effectively process regions with activity and limits performance near the noise floor.

### B. Effect of IFC parameters

The effect of IFC threshold on the performance is studied across the regions of analysis. While the PSNR at regions C and D decreases gradually as the IFC threshold increases, the PSNR near the noise floor is consistently less than 20dB. Moreover, as the IFC threshold increases, the mean pulse rate decreases exponentially. From Fig. 3, it is evident that proper selection of IFC threshold guarantees lower pulse rates without degrading performance.

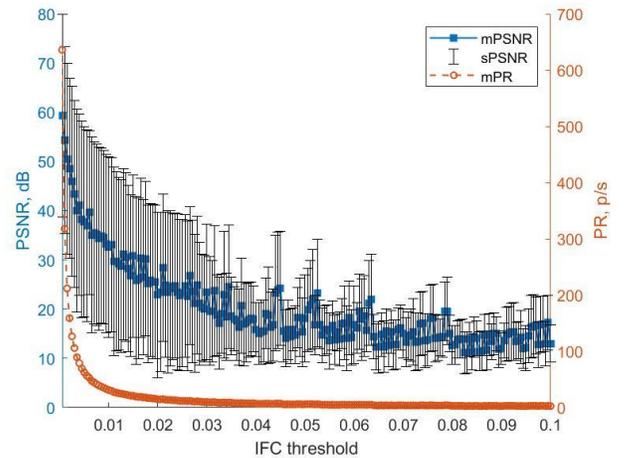

Fig. 3. Effect of IFC threshold on performance. mPSNR and sPSNR denote the mean and standard deviation of PSNR respectively, and mPR denotes the mean of pulse rate.

The effect of IFC rate of decay and approximations to the algorithm is presented in Fig. 4. Comparison of performance of algorithm with and without approximation reveals PSNR at regions A and B to be similar in both cases while PSNR at regions C and D are significantly different (p<0.05) with the approximated algorithm having lower mean PSNR as shown in Fig. 4a. Moreover, the mean pulse rate of the algorithm with and without approximation across variations in IFC rate of decay is significantly different (p<0.05) as shown in Fig. 4b, with approximated algorithm having lower pulse rate. While the approximated algorithm in Table II offers simpler implementation dependent only on pulse intervals and provides sparse representation at higher rates of decay, there is trade-off in PSNR at high amplitude regions when compared with the algorithm in Table I.

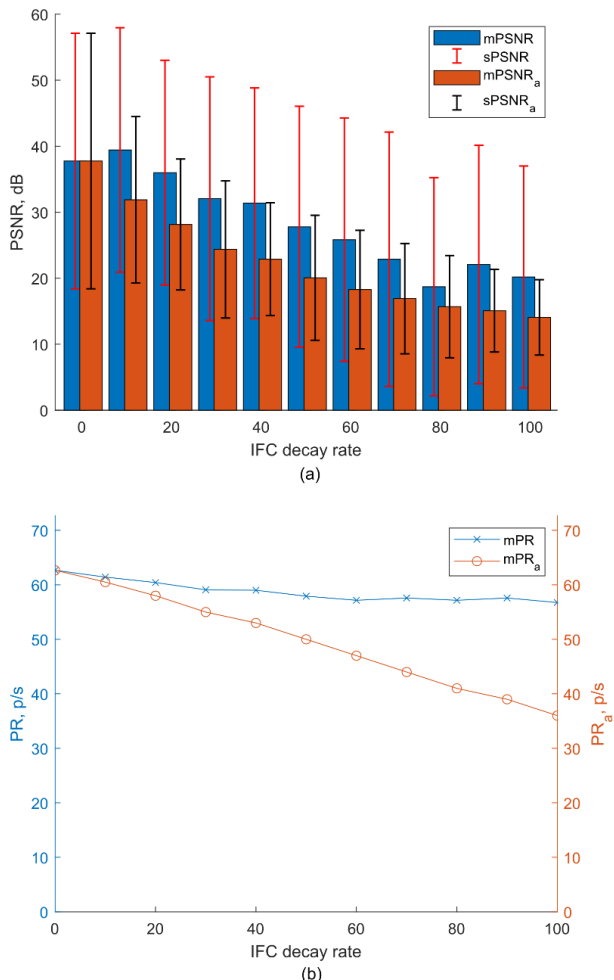

Fig. 4. Effect of IFC rate of decay and approximations on performance. The subscript 'a' in legend denotes the approximated algorithms.

### C. Comparison with digital processing

In Fig. 5, the performance curves for the comparison of the proposed algorithm with digital processing is presented. The performance is similar to Fig. 3, and PSNR is proportional to the signal amplitude for a given IFC threshold. It is to be noted that exact reconstruction for IFC is impossible and these results are obtained with the approximate method developed by Feichtinger et al. [15]. While processing of reconstructed signals from pulse trains is not the focus of the paper, substantially higher pulse rates than corresponding Nyquist rates is required for applications that necessitates high fidelity after signal reconstruction. The above behavior of the performance curves also holds true for pulse train multiplication and convolution.

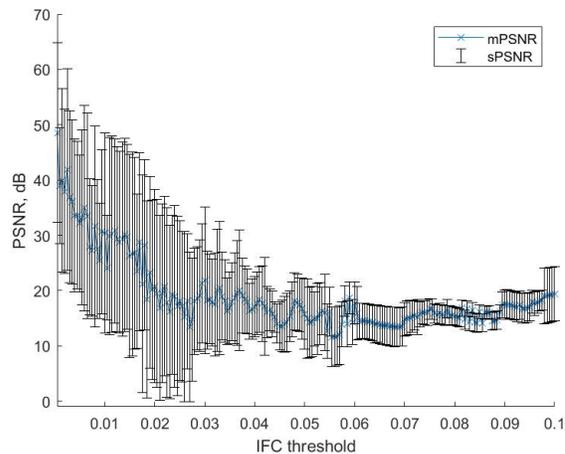

Fig. 5. Comparison of proposed algorithm with digital processing after reconstruction of operand pulse trains.

### D. Semantic information processing in ECG

The processing and representation of relevant information in ECG signal using pulse trains is demonstrated in Fig. 6. The top panel of Fig. 6 shows the ECG signal with and without baseline wander. Ideally, in the absence of baseline noise, the cardiac events are selectively captured in the pulse representation of ECG signal and the isoelectric deviations with no activity are not represented [13]. However, during baseline deviations, pulses corresponding to cardiac events are obfuscated by higher pulse density due to shift in baseline as shown in the middle panel of Fig. 6.

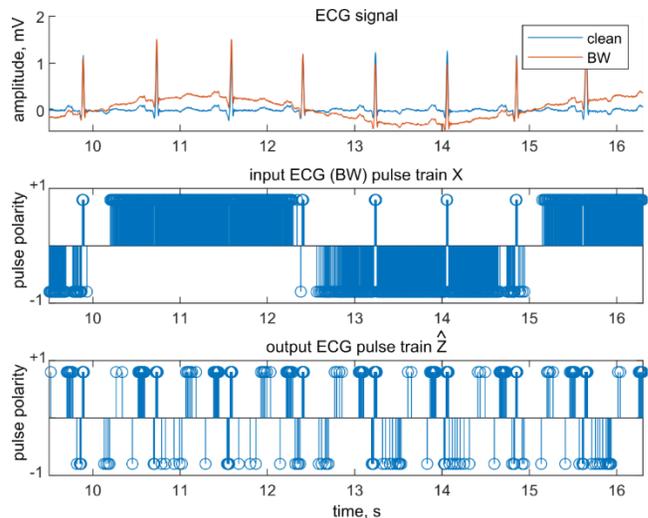

Fig. 6. Semantic processing of ECG signal using pulse trains. 'BW' denotes baseline wander in the signal.

With the proposed algorithm, pulses are directly processed to subtract the baseline noise as shown in the bottom panel of Fig. 6 and boundaries of relevant features of interest such as P wave, QRS complex, and T wave are clearly delineated in the pulse representation. Moreover, the pulse rate of ECG signal is reduced from 61.25 ± 100.57 p/s to 15.3 ± 10.5 p/s after processing, with PSNR and $r$ of 20.12 ± 8.76 dB and 0.73 ± 0.32 respectively. Thus, processing signals using pulses highlights the relevant information content with sparse representation.

## VII. DISCUSSION

Representation and processing of semantic information in signals is critical for mobile wireless sensor networks and IoT applications [20]. Prior research has shown that the IFC pulse conversion enable sparse representation of features of interest in physiological signals such as ECG [12], [13], neural data [37], and photoplethysmogram [14]. This article presented algorithms for processing the pulse trains created by IFC directly without signal reconstruction, and demonstrated processing of the semantic information in ECG signal.

Simulations with synthetic data show that pulse based signal processing has PSNR proportional to signal amplitude, with limited pulse representation near noise floor. Precision of the pulse based operations is not uniform over the dynamic range of the signal amplitude, which is desirable in long term monitoring applications to represent semantic information with high precision while suppressing background noise.

In digital signal processing, to meet desired performance specifications, the number of samples to which the digital data must be interpolated is determined, and then the operations are performed sample-by-sample. However, in pulse signal processing, the threshold of IFC that satisfies the performance criteria is determined, and then the operations are done on the areas between consecutive pulses. Feichtinger et al. [15] showed that bandlimited functions are not completely determined by the IFC and the same results hold true for processing of multiple operands i.e., there are non-zero bandlimited signals that will never produce pulses at the output even though the input operands have pulses. The trade-offs in terms of accuracy for the IFC parameters is studied in [37] and operating ranges are selected based on specifications to provide the right balance between performance and sparseness.

Hardware implementations of the proposed algorithms require digitization of the time axis. Moreover, implementing pulse signal processing in hardware requires new approaches as operations need to be performed on area between pulses that occur non-uniformly. Recently, Nallathambi et al. [41], implemented 16-bit pulse adder based on the approximated algorithms in Table II. Their system, synthesized in SMIC 0.18μm (100MHz) CMOS process, consists of quantization clock and time counters as building blocks of pulse-based arithmetic. The preliminary results demonstrate the feasibility of signal processing with pulse trains in hardware.

The alternative approach to the present work is to reconstruct signals from pulses, perform digital processing and convert the processed signals to pulses. In such scenarios, the accuracy of the operations will be limited by the approximate reconstruction procedure [15], and increasing the fidelity requires reduction of threshold substantially, thereby impacting the data rates. By processing the pulses directly with the proposed algorithm, we circumvent both the complexity of the signal reconstruction algorithm and the subsequent process of IFC conversion. With the proposed algorithms, the focus is on applications where reconstruction is not necessarily the goal but tasks such as classification and anomaly detection that require representation of semantic features in signal.

In this article, an example of processing semantic information in ECG signals using pulse trains is presented. The mean pulse rate of the ECG is less than 20p/s, which is drastically lower than existing IoT-based cardiac patient monitors that require at least 125 samples per second [42]. In general, pulse trains are well suited for processing semantic information in transient signals such as ECG, EEG, seismological recordings, radar and others, which are embedded in noisy backgrounds.

## VIII. CONCLUSION

The present work provides an alternative to conventional digital signal processing techniques for performing arithmetic operations on continuous amplitude and time signals using pulse trains generated from IFC. The proposed algorithms enable online implementation of the theoretical framework for pulse-based computation. The results with synthetic and natural data demonstrate the capability of the algorithms in processing semantic information content in the signals using pulse trains.